\documentclass[12pt]{article}
\usepackage{latexsym,graphicx}

\newcommand{\sbar}{\hspace{0.2em}\overline{\rule[0.42em]{0.4em}{0em}}
\hspace{-0.5em}s\hspace{0.1em}}
\newcommand{\rs}{\langle r^2\rangle\rule[-0.2em]{0em}{0em}_s}

\begin{document}
\thispagestyle{empty}
\begin{center}
{\Large\bf  Bounds on the slope and the curvature
\vskip 0.1cm
of the scalar $K\pi$ form factor \\
\vskip 0.3cm
at zero momentum transfer}
\vskip1.4cm
C. Bourrely $^a$ and  I. Caprini $^b$
\vskip0.3cm
$^a$ Centre de Physique Th\'eorique, UMR 6207,\\ CNRS-Luminy, Case 907, \\
F-13288 Marseille Cedex 9 - France
\vskip 0.2cm
$^b$ National Institute of Physics and Nuclear Engineering,\\
POB MG 6, Bucharest, R-077125 Romania
\vskip 2cm
\end{center}
\begin{abstract}
We derive and calculate unitarity bounds on the slope and curvature of the
strangeness-changing scalar form factor at zero momentum transfer, 
using low-energy constraints and Watson final state interaction theorem.
The results indicate that the curvature is important and should not be
neglected in the representation of experimental data.
The bounds can be converted also into an allowed region for the constants
$C_{12}^r$ and  $C_{34}^r$ of Chiral Perturbation Theory.
Our results are consistent with, but weaker than the  predictions
made by Jamin, Oller and Pich  in a coupled channel dispersion  approach based
on chiral resonance model.
We comment on the differences between the two dispersive
methods and argue that the unitarity bounds are useful as an independent
check involving different sources of information.
\end{abstract}
\vskip 2cm
\noindent CPT-2005/P.017
\newpage
\section{Introduction}
The scalar form factor relevant for the decay $K\rightarrow \pi\ell\nu$ is
proportional to the matrix element $\langle K|\sbar u|\pi\rangle$. It is
 related to the vector form factors $f_\pm(t)$ by the expression
\begin{equation}\label{f}
f_0(t)=f_+(t) +\frac{t}{M_K^2-M_\pi^2}\, f_-(t)\,,
\end{equation}
in  the standard normalization where $f_0(0)$ coincides with $f_+(0)$, a
quantity that is of central
importance for the determination of the CKM matrix element $V_{us}$.
Writing the Taylor expansion of the form factor near $t = 0$
\begin{equation}\label{Taylor}
f_0(t)=f_+(0)\left\{1+\lambda_0 \,t +c \,t^2+\ldots\right\}\,,
\end{equation}
where $\lambda_0$ is the slope and $c$ the curvature, it is customary to
relate the slope with the scalar radius of the pion, through
$\lambda_0\equiv\rs^{K\pi} M_\pi^2/6$. This expansion is usually considered
in the analysis of experimental data on
$K_{l3}$ decays of both charged and neutral kaons.
The experimental situation on the $K_{l3}$ decays improved very much recently.
For the neutral kaons, the new result from KTeV collaboration is 
$\langle r^2\rangle\hspace{-0.1em}
\mbox{\raisebox{-0.3em}{$\stackrel{K_L \pi}{\hspace{-1em}
\mbox{\footnotesize\it s}}$}}= 0.165 \pm 0.016\,\mbox{fm}^2$
\cite{KTeV_form_factors_2004}. This value  now dominates the
statistics and  is consistent with the 1974 high statistics
experiment \cite{Donaldson_1974}. For the charged kaons, the data collected
at the ISTRA+ detector \cite{Yushchenko_2003} give the radius
$\rs^{K^\pm\pi}=0.235\pm 0.014\pm 0.007\,\mbox{fm}^2$, 
which  dominates the world average.

On the theoretical side there is also a considerable progress in the study of
the scalar strangeness changing form factor $f_0(t)$. Low energy theorems 
impose strong constraints on the values of this function at some particular 
points. The value at the origin $f_0(0)\approx 1$ is given by Ademollo-Gatto
relation \cite{AdGa}, while the theorem of Callan and Treiman, refined by 
Dashen and Weinstein \cite{Dashen_Weinstein}, states that in the limit of 
vanishing quark masses the value of the form factor at the special point 
$t=M_K^2-M_\pi^2$ is
\begin{equation}\label{CT}
f_0(\Delta_{K\pi})= F_K/F_\pi +\Delta_{CT},\quad\quad\quad\quad
\Delta_{K\pi}=M_K^2-M_\pi^2,
\end{equation}
where the $O(\hat m)$ correction $\Delta_{CT}$ was calculated to one loop in
Chiral Perturbation Theory ($\chi$PT) \cite{GL_form_factors}
and found to be tiny: $\Delta_{CT}\approx - 3\cdot 10^{-3}$.  
Also,  in Ref. \cite{GL_form_factors} a first prediction to one loop for the 
radius was obtained,
$\rs^{K\pi}=0.20\pm 0.05\,\mbox{fm}^2$. 
It is shown that the corrections of $O(\hat{m})$ do
not contain a chiral logarithm of the type $M_\pi^2\log M_\pi^2$ and therefore
they are very small \cite{GL_form_factors}.
Recently, the $K_{\ell 3}$ form factors were calculated to two loops of 
$\chi$PT \cite{Post_Schilcher_2002,Bijnens_Talavera_2003}.

Dispersion theory provides another useful tool for making predictions on the
form factors  that govern the $K_{l3}$ decay.
These functions are  analytic in the $t$-plane cut from the unitarity threshold
$(M_K+M_\pi)^2$ to infinity.
A first approach to derive bounds on the values of the form factors and their
derivatives in the physical region was proposed in the years 70. It is
based on analyticity, and exploited
the K\"allen-Lehmann representation for a suitable two-point function
$\psi(q^2)$, in combination with unitarity and low-energy theorems.
However, these first studies \cite{Okubo}-\cite{AuMa}
were based on an unsubtracted dispersion relation for $\psi(q^2)$, which 
turned out to be incorrect: this function,
calculated in perturbative QCD \cite{BNRY}, satisfies a dispersion relation
which requires two subtractions.
The correct dispersion relation was applied for the first time in Ref.
\cite{BoMaRa}. Instead of using the low energy theorems on $\psi(0)$ as input
\cite{Okubo}-\cite{AuMa}, the new approach involves the 
calculation of the
derivative $\psi''(Q^2)$ obtained  by perturbative QCD in the deep euclidian
region. The constraining power of this condition is weaker and therefore
the strength of the bounds is reduced.

An alternative dispersive approach based on coupled channel unitarity 
equations for the 
$K\pi$, $K\eta$ and $K\eta'$ form factors was considered
recently in \cite{Jamin_Oller_Pich_2002}. It is based on unitarized 
chiral perturbation plus a $K$-matrix fit of the S-wave 
scattering \cite{Jamin_Oller_Pich_2000}. 
In a recent paper \cite{Jamin_Oller_Pich_2004}, the authors combine the values
of the slope and curvature
of the scalar form factor at zero momentum transfer, calculated in
\cite{Jamin_Oller_Pich_2002}, with the two-loop calculation of $\chi$PT
\cite{Post_Schilcher_2002,Bijnens_Talavera_2003}, obtaining thereby predictions
for the chiral constants $C_{12}^r$ and  $C_{34}^r$.

In the present work we revisit and extend the technique of unitarity 
bounds with the aim
of deriving constraints on the slope and the curvature appearing in the Taylor
expansion (\ref{Taylor}). The study is motivated by the  recent interest in
the determination of these parameters, both experimentally and in $\chi$PT. On
the other hand, since the last work on unitarity bounds \cite{BoMaRa},
the knowledge of the quantities entering as input improved considerably. 
For instance, the correlator $\psi(q^2)$ is now  calculated
in perturbative QCD to order $\alpha_s^3$
\cite{GoKa}, \cite{Chet}, and recent lattice calculations of the 
light and strange
quark masses \cite{QCDSF}, \cite{MILC1} and the ratio $F_K/F_\pi$ \cite{MILC2}
are available. We notice also that the implementation of  Watson final state
interaction theorem \cite{Watson} was never done with the correct dispersion
relation for $\psi(q^2)$ (in Refs. \cite{Bour}-\cite{AuMa}, where this
problem was considered, the old unsubtracted dispersion relation was used). 
In the present work we implement simultaneously both low energy constraints 
and the Watson theorem, exploting in an optimal way the formalism,
which can be compared with the alternative study made in
Ref. \cite{Jamin_Oller_Pich_2004}.

In Section 2, we give a short review of the mathematical technique we will use,
and show  how to incorporate additional constraints provided by $\chi$PT  and
Watson final state interaction theorem.  
In Section 3,  we discuss our results and compare them
with the recent predictions on the slope and curvature of the scalar $K\pi$
form factor presented in 
\cite{Jamin_Oller_Pich_2002}-\cite{Jamin_Oller_Pich_2004}. As in these 
references we work in the
isopsin limit (the isospin breaking corrections calculated in \cite{RadCor} 
are important and should be included in a more precise analysis). 
We end the paper with some comments on the usefulness of the method and its
possible generalizations.

\section{Unitarity bounds}
The starting point in the derivation of the bounds is the two-point function
\begin{equation}\label{psi}
\psi(q^2)=i\int {\rm d}^4 x e^{i q\cdot x} \langle 0|{\rm T} (\partial_\mu
V^\mu (x) \partial_\mu  V^\mu (0)^\dagger)|0\rangle\,,
\end{equation}
where $V_\mu=\sbar \gamma_\mu u$ is the strangeness changing vector current
that governs $K_{l3}$ decay. 
The function $\psi(q^2)$ was calculated for euclidian arguments
$Q^2=-q^2 >0$ up to corrections of order $\alpha_s^3$ \cite{BNRY}, \cite{GoKa},
\cite{Chet}. It grows asymptotically as $Q^2$ and satisfies
a twice-subtracted  K\"allen-Lehmann representation. One can use either the
dispersion relation for the second derivative  $\psi^{''}(Q^2)$
\begin{equation}\label{drpsi}
 \psi^{''}(Q^2)=\frac{2}{\pi}\int\limits_{t_+}^\infty {\rm d}t
\frac{{\rm Im} \,\psi(t)}{ (t+Q^2)^3}\,,
\end{equation}
where  $t_+ =(M_K + M_\pi)^2$ is the first unitarity threshold,
or the relation satisfied by the derivative of the ratio $\psi(Q^2)/ Q^2$:
\begin{equation}\label{drpsi1} 
\left(\frac{\psi(Q^2)}{Q^2}\right)' +
\frac{\psi(0)}{Q^4}=\frac{1}{\pi}\int\limits_{t_+}^\infty {\rm d}t
\frac{{\rm Im}\, \psi(t)}{ t(t+Q^2)^2}\,.
\end{equation}
The advantage of the last relation is that it incorporates the value $\psi(0)$,
which satisfies the inequality
\begin{equation}\label{psi0}
\psi(0)< (M_K F_K - M_\pi F_\pi)^2 \,,
\end{equation}
derived by Mathur and Okubo \cite{MaOk} (we use the normalisation  $F_\pi=
92.4$ MeV). In our analysis we shall use the relations (\ref{drpsi}) and
(\ref{drpsi1}) successively.

Unitarity bounds  \cite{Okubo} are obtained by combining the above dispersion
relations with the inequality
\begin{equation}\label{unit}
{\rm Im}\, \psi(t)\ge \frac{3  (M_K^2-M_\pi^2)^2 }{64 \pi}\frac{\sqrt{(t-t_+)
(t-t_-)}}{t}\, |f_0(t)|^2,  \quad t_\pm =(M_K\pm M_\pi)^2,
\end{equation}
given by unitarity and the positivity of the spectral function.
Then Eq. (\ref{drpsi}) gives \cite{BoMaRa}
\begin{equation}\label{L2}
\frac{1}{\pi}\int\limits_{t_+}^\infty {\rm d}t\, w(t) |f_0(t)|^2
\le \psi^{''}(Q^2),
\end{equation}
where \begin{equation}\label{w}
w(t)= {3 (M_K^2-M_\pi^2)^2 \over 32 \pi}\,{\sqrt{(t-t_+) (t-t_-)}\over
t(t+Q^2)^3}\,.
\end{equation}
Similarly  Eq.(\ref{drpsi1}) leads to
\begin{equation}\label{L21}
\frac{1}{\pi}\int\limits_{t_+}^\infty {\rm d}t\, \tilde w(t) |f_0(t)|^2
\le  \left(\frac{\psi(Q^2)}{Q^2}\right)' + \frac{\psi(0)}{Q^4} ,
\end{equation}
where \begin{equation}\label{tildew}
\tilde{w}(t)= w(t) \,\frac{t+Q^2}{2t}\,.
\end{equation}
The  r.h.s. of (\ref{L2}) is given by perturbative QCD
\cite{BNRY},\cite{Chet}
\begin{eqnarray}\label{QCD}
&&\psi^{''}(Q^2) = \frac{3}{8\pi^2}\frac{( m_s(Q^2) -  m_u(Q^2))^2}
{Q^2} \times \nonumber \\
&&\left[1 + \frac{11}{3}\,\frac{ \alpha_s(Q^2)}{\pi}  +14.17
\left(\frac{\alpha_s(Q^2)}{\pi}
\right)^2+\ldots+   O\left(\frac{ m^2}{Q^2}\right) +O\left(\frac{1}{Q^4}\right)
\right],
\label{psi2nd}
\end{eqnarray}
while the corresponding expansion for the r.h.s. of (\ref{L21}) reads 
\cite{GoKa}
\begin{eqnarray}\label{QCD1}
&&\left(\frac{\psi(Q^2)}{Q^2}\right)' = \frac{3}{8\pi^2}\frac{( m_s(Q^2) -
m_u(Q^2))^2}{Q^2} \times\nonumber \\
&&\left[1 + \frac{17}{3}\,\frac{ \alpha_s(Q^2)}{\pi}  + 45.84
\left(\frac{\alpha_s(Q^2)}{\pi}
\right)^2+\ldots+   O\left(\frac{ m^2}{Q^2}\right) +O\left(\frac{1}{Q^4}\right)
\right]\,.
\label{psi1}
\end{eqnarray}
Using standard techniques \cite{Okubo}-\cite{BoMaRa}, \cite{Ca}, the conditions
(\ref{L2}) or (\ref{L21}) can be converted into bounds on
the values of  $f_0(t)$ for points inside the analyticity domain,
in particular, on the slope  
and the curvature at the origin. We consider in detail the condition
(\ref{L2}), indicating at the end of the section the modifications
resulting for the condition (\ref{L21}). 

The problem is brought into a canonical form by
making the conformal mapping
\begin{equation}\label{z}
z(t)={\sqrt{t_+}-\sqrt{t_+-t}\over \sqrt{t_+}+\sqrt{t_+-t}}\,,
\end{equation}
which maps the $t$-plane cut for $t>t_+$ onto the disk $|z|<1$, such that
$z(0)=0$, $z(t_+)=1$, $z(\infty)=-1$,  and
the upper edge of the cut $t>t_+$ becomes the upper semicircle  $z=\exp (i
\theta)$, $0<\theta<\pi$. The inverse of (\ref{z}) and its derivative are
\begin{equation}\label{tz}
t(z)= 4 t_+\, z/(1+z)^2\,,\quad  {\rm d} t/ {\rm d}z = 4 t_+ (1-z)/ (1+z)^3,
\end{equation}
while for points on the boundary, $z=e^{i \theta}$, Eqs. (\ref{z}) 
and (\ref{tz}) write
\begin{equation}\label{theta}
\theta=2 \arctan \sqrt{t/ t_+-1}\,,\quad t(e^{i\theta})=t_+/\cos^2{1\over 2}
\theta\,.
\end{equation}
In the new variable the relation
(\ref{L2}) becomes
\begin{equation}\label{L2z}
{1\over 2\pi}\int\limits_{-\pi}^{\pi} {\rm d} \theta \rho(\theta)
|f(e^{i\theta})|^2 \le   \psi^{''}(Q^2),
\end{equation}
where $f(z)=f_0(t(z))$ and
\begin{equation}\label{rho}
\rho(\theta)=   w(t(e^{i\theta})) \,\left\vert{{\rm d} t(e^{i\theta})\over 
{\rm d} \theta} \right\vert\,.
\end{equation}
We define the outer function $C(z)$, analytic and without zeros inside
$|z|<1$, such that on the boundary of the unit disk:
\begin{equation}\label{Crho}
|C(e^{i\theta})|= \sqrt{\rho(\theta)}\,.
\end{equation}
A straightforward calculation \cite{AuMa}, \cite{Ca} gives
\begin{equation}\label{C}
C(z)=\frac{1}{4}
{\sqrt{3\over 2\pi}}\, { M_K-M_\pi \over M_K +M_\pi} \, (1-z) (1+z)^{3/2}
\frac{\sqrt{1-z+\beta (1+z)}}{(1-z+\beta_Q (1+z))^3} \,,
\end{equation}
where $\beta =\sqrt{1-t_-/t_+}$ and $\beta_Q =\sqrt{1+Q^2/t_+}$.
Then the inequality (\ref{L2z}) can be written in the form
\begin{equation}\label{L2g}
\frac{1}{2\pi}\int\limits_{-\pi}^{\pi} {\rm d} \theta\,  |g(e^{i\theta})|^2
\le \psi^{''}(Q^2)\,,
\end{equation}
where the function $g(z)$ defined by
\begin{equation}\label{g}
g(z)= C(z)\, f(z)
\end{equation}
is analytic  inside the disk $|z|<1$. Expanding this function as
\begin{equation}\label{gTaylor}
g(z)= \sum\limits_{n=0}^\infty g_n z^n,
\end{equation} we write the $L^2$ norm condition (\ref{L2g}) in terms of the 
Taylor coefficients $g_n$ as
\begin{equation}\label{L2gn}
\sum\limits_{n=0}^\infty g_n^2\le   \psi^{''}(Q^2) \,.
\end{equation}
We mention that the reality condition $f_0(t^*)=
f^*_0(t)$ means that the coefficients $g_n$ are real.

From Eq. (\ref{g}) it follows that each Taylor coefficient $g_n$ is a
linear expression of the first $n$ derivatives
of the form factor $f_0(t)$ with respect to $t$ at  $t=0$
(we recall that  $t=0$  is transformed  by (\ref{z})
into the origin $z=0$), in particular $g_0$, $g_1$ and $g_2$ are expressed in
terms of the parameters $f_0(0)$, $\rs^{K\pi}$ and $c$.
The coefficients depend on  the conformal mapping $z(t)$ given in (\ref{z}) 
and the  
outer function  $C(z)$ given in (\ref{C}), where $Q^2$ enters as a parameter.
For instance, for $Q^2= 4\, {\rm GeV}^2$, a straightforward calculation gives:
\begin{eqnarray}
g_0&=&   f_0(0)\, 0.00164 \,,\nonumber \\
g_1&=& f_0(0)\, [-0.00189 + 0.0113 \,\rs^{K\pi}\,]\, , \label{rel}\\
g_2&=& f_0(0) [-0.000259 - 0.03579 \rs^{K\pi} + 0.00428\, c\, ]\,. \nonumber
\end{eqnarray}
In order to find the maximum domain allowed
for the  parameters $f_0(0)$, $\rs^{K\pi}$ and $c$,  we fix the first three
coefficients $g_0$, $g_1$ and $g_2$, and  take the minimum of the left hand
side of (\ref{L2gn}) with respect to the remaining parameters $g_n$, $n\ge 3$,
which are arbitrary. Clearly, the minimum is reached for $g_n=0$, $n\ge 3$, so
the allowed domain becomes
\begin{equation}\label{g0g1g2}
g_0^2 + g_1^2+g_2^2 \le     \psi^{''}(Q^2) \,.
\end{equation}
Using Eq. (\ref{rel}), this inequality defines a convex domain for  the
parameters $f_0(0)$, $\rs^{K\pi}$ and  $c$. For a fixed $f_0(0)$, this domain
is the interior of an ellipse in the plane ($\rs^{K\pi},\, c$).

The strength of the bounds, expressed by the size of the allowed domain,
is increased by imposing additional contraints.
\vskip 0.3 cm
\noindent {\it (1) The Callan-Treiman condition}

Let us consider first the Callan-Treiman condition (\ref{CT}). We note by
$z_0$ the image of the point $t_0=\Delta_{K\pi}$ in the $z$-plane:
$z_0=\frac{\sqrt{M_K/M_\pi+1}-\sqrt{2}}
{\sqrt{M_K/M_\pi+1}+\sqrt{2}}$. From Eqs. (\ref{g}) and (\ref{gTaylor}) we 
have the condition
\begin{equation}\label{CTg}
\sum\limits_{n=0}^\infty g_n z_0^n = C(z_0) f_0(\Delta_{K\pi})\,.
\end{equation} As explained in Ref. \cite{Ca}, in order to obtain the allowed 
domain, we have to evaluate the minimum of the left hand side of (\ref{L2gn}) 
with respect to the coefficients $g_n$, for $n\ge 3$
subject to the condition (\ref{CTg}). We impose this condition by the Lagrange
multipliers method. The Lagrangian is
\begin{equation}\label{L}
{\mathcal L} = \sum\limits_{n=0}^\infty g_n^2\,+\, \mu
\,\left(\sum\limits_{n=0}^\infty g_n z_0^n- C(z_0) f_0(\Delta_{K\pi})  \right)
\end{equation}
where $\mu$ is the Lagrange multiplier. The minimizing condition
\begin{equation}\label{eqs}
{\partial {\mathcal L}\over \partial g_n}=0\,, \quad n\ge 3\,,
\end{equation}
has the solution $g_n=-\mu z_0^n/2$ for  $n\ge 3$. Inserting this solution
in the condition (\ref{CTg}), we find  the Lagrange multiplier
\begin{equation}\ref{mu}
\mu = 2 \,{1-z_0^2\over z_0^6}\,\left(g_0+g_1 z_0+g_2 z_0^2 -C(z_0)
f_0(\Delta_{K\pi})\right) \,.
\end{equation}
By inserting the solution of Eqs. (\ref{eqs}) with $\mu$ from (\ref{mu}) into 
the left hand side of Eq. (\ref{L2gn}), we obtain the inequality
\begin{equation}\label{g0g1g2CT}
g_0^2 + g_1^2+g_2^2 + {1-z_0^2\over z_0^6} \,\left(g_0+g_1 z_0+g_2 z_0^2
-C(z_0) f_0(\Delta_{K\pi})\right)^2 \,\le  \psi^{''}(Q^2)\,,
\end{equation}
which represents again, for each $f_0(0)$, the interior of an ellipse in
the plane $(\rs^{\pi K}, ~c)$.\\
\vskip 0.3cm
\noindent {\it (2) The Callan-Treiman condition and the Watson theorem}

We now include Watson theorem \cite{Watson}, which states that
\begin{equation}\label{Watson}
{\rm Arg}[f_0(t+i\epsilon)] = \delta_0^{1/2}(t)\,, \quad (M_K+M_\pi)^2 <t <
t_{in}\,,
\end{equation}
where $\delta_0^{1/2}(t)$ is the $I=1/2$ phase shift of the $S$ wave $K\pi$ 
elastic scattering and $t_{in}=(M_K+ 3M_\pi)^2$
is the inelastic threshold. 
The incorporation of Watson theorem in the unitarity bounds for the $K_{l3}$ 
form factors was done in an approximate way in Ref. \cite{Bour}
and rigorously in Ref. \cite{AuMa}.
However, as mentioned above, in these works  it was assumed that
$\psi(q^2)$ satisfies  an unsubtracted dispersion relation,
which was not confirmed in perturbative QCD. Moreover, the authors derived
bounds on the magnitude $|f_0(t)|$ in the physical region, and did not 
consider  the slope and the curvature in which we are now interested.
In the present paper, we apply a mathematical method based on Lagrange
multipliers, which is equivalent to the technique adopted in \cite{AuMa} but 
is conceptually simpler.  
The details are explained in Ref. \cite{Ca}, where the
allowed domain for the coefficients $g_n$ with the constraint (\ref{Watson})
was obtained.  Below, we consider the complete problem by imposing 
simultaneously the conditions (\ref{CT}) and (\ref{Watson}).

Let us note by  $\theta_{in}= 2 \arctan \sqrt{t_{in}/t_+-1}$ the image of the
point $t_{in}$ on the boundary of the unit disk $|z|<1$ in the $z$-plane.
We define the Omn\`es function \cite{AuMa}, \cite{Omnes}
\begin{equation}\label{Omnes}
O(z)={\rm exp} \frac{i}{\pi} \int\limits_{-\pi}^\pi {\rm d}\theta
\frac{\delta(\theta)}{1-z e^{-i\theta}}\,,
\end{equation}
where $\delta(\theta)=\delta_0^{1/2}(t(e^{i\theta}))$ for
$0<\theta<\theta_{in}$. When $\theta>\theta_{in}$ the function $\delta(\theta)$
is arbitrary, however these values will not enter the result.
We recall again  the reality condition $f_0(t^*)=
f^*_0(t)$, which  means that $\delta(\theta)$ is an odd function of $\theta$:
$\delta(-\theta)=-\delta(\theta)$. As in Ref. \cite{AuMa}, we assume that
$\delta(\theta)$ is a Lipschitz continuous function of order larger 
than 1/2 on $[0,\,\theta_{in}]$. By construction, $O(z)$ is analytic and 
without
zeros inside $|z|<1$, and ${\rm Arg}\, O(e^{i\theta})=\delta(\theta)$. 
Therefore, in the ratio $f(z)/O(z)$  the phases of the numerator and 
the denominator compensate, leading to a real quantity along the 
elastic region.
Expressed in terms of the function $g(z)$ defined in (\ref{g}) and its Taylor
coefficients $g_n$, this condition reads
\begin{equation}\label{gW}
\lim_{r\to 1}{\rm Im}\left[
\frac{g(re^{i\theta})}{W(re^{i\theta})}\right]=\lim_{r\to
1}\sum\limits_{n=0}^\infty g_n\, {\rm Im} \left[\frac{r^n
e^{in\theta}}{W(re^{i\theta})}\right] =0\,, \quad
-\theta_{in}<\theta<\theta_{in}\,,
\end{equation}
where we introduced the function
\begin{equation}\label{W}
W(z)= C(z) \,O(z)\,.
\end{equation}
We must find the minimum of the left hand side of (\ref{L2gn}) with the
constraints  (\ref{CTg}) and (\ref{gW}).
We consider the Lagrangian \cite{Ca}
\begin{eqnarray}\label{L1}
{\mathcal L}& =& \sum\limits_{n=0}^\infty g_n^2\,+\, \mu
\,\left(\sum\limits_{n=0}^\infty g_n z_0^n- C(z_0)
f_0(\Delta_{K\pi})\right)\nonumber\\
&&+\lim_{r\to 1}\sum\limits_{n=0}^\infty g_n
\,\int\limits_{-\theta_{in}}^{\theta_{in}} \frac{2 {\rm d}\theta'} {\pi}
\lambda(\theta') |W(e^{i\theta'})|\, {\rm Im} \left[\frac{r^n
e^{in\theta'}}{W(re^{i\theta'})}\right],
\end{eqnarray}
where $\lambda(\theta)$ is a generalized Lagrange multiplier, which must be an
odd function:  $\lambda(-\theta)=-\lambda(\theta)$.  
As discussed in Ref. \cite{Ca}, the factor $|W(e^{i\theta})|$  was inserted
in the integrand in order to have a function $\lambda(\theta)$ with integrable
squared modulus ({\em i.e.} of class  $L^2$), while the 
coefficient $2/\pi$ is introduced for convenience.

The equations (\ref{eqs}) give the optimal coefficients $g_n$:
\begin{equation}\label{gn}
g_n=-\frac{\mu z_0}{2}- \lim_{r\to
1}\int\limits_{-\theta_{in}}^{\theta_{in}}\frac {{\rm d}\theta'}{\pi}
\lambda(\theta') |W(e^{i\theta'})|\, {\rm Im} \left[\frac{r^n
e^{in\theta'}}{W(r e^{i\theta'})}\right],\quad n\ge 3\,.
\end{equation}
By imposing the constraint (\ref{CTg}) we find the Lagrange multiplier $\mu$
\begin{equation}\label{mu}
\mu=  \frac{2(1-z_0^2)}{ z_0^6} \,\left[g_0+g_1 z_0+g_2 z_0^2 -C(z_0)
f_0(\Delta_{K\pi}) -z_0^3   \int\limits_{-\theta_{in}}^{\theta_{in}}\frac{
{\rm d}\theta'}{\pi} \lambda(\theta') h(\theta')\right]\,,
\end{equation}
where we used the notation
\begin{equation}\label{h}
h(\theta)={\rm Im}\,\left[\frac{e^{3i\theta-i \Phi}}{1-z_0
e^{i\theta}}\right]=\frac{\sin[3\theta-\Phi(\theta)]-z_0
\sin[2\theta-\Phi(\theta)]}{1+z_0^2-2 z_0\cos\theta}\,,
\end{equation}
with
\begin{equation}\label{Phi}
\Phi(\theta)= {\rm Arg}\, C(e^{i\theta})+\delta(\theta)\,.
\end{equation}
This function  is readily obtained using the expression (\ref{C}) of the outer
function $C(z)$
\begin{equation}\label{Phi1}
\Phi(\theta)= \mp \frac{\pi}{2}+ \frac{5}{4}\theta +\frac{1}{2}{\rm
Arg}[1-e^{i\theta}+\beta (1+e^{i\theta})] - 3 {\rm Arg}[1-e^{i\theta}+\beta_Q
(1+e^{i\theta})] +\delta(\theta)\,,
\end{equation}
the signs of the first term corespond to $\theta>0$ and $\theta<0$,
respectively. By including into the condition (\ref{gW}),
the optimal coefficients  (\ref{gn}), where $\mu$ is
given in (\ref{mu}), and using the Plemelj-Privalov relation \cite{Musk}
\begin{equation}\label{PP}
\lim_{r\to 1}\frac{1}{\pi}\, \int\limits_{-\pi}^{\pi} {\rm d}\, \theta'
\frac{F(\theta')}{1-r\, e^{i(\theta-\theta')}}= F(\theta) + \frac{1}{\pi}\,
\int\limits_{-\pi}^{\pi} {\rm d}\, \theta' \frac{F(\theta')}{1-
e^{i(\theta-\theta')}}\,,
\end{equation}
where the last integral is a Principal Value, we obtain an integral
equation  for the function $\lambda(\theta)$ given by
\begin{eqnarray}\label{inteqCTW}
V(\theta)=\lambda(\theta)&&-\int\limits_{-\theta_{in}}^{\theta_{in}}
\frac{ {\rm d}\theta'} {2\pi}
\lambda(\theta')\,\frac{\sin[5/2(\theta-\theta')-\Phi(\theta)+\Phi(\theta')]}
{\sin (\theta-\theta')/2}\nonumber\\
&& -(1-z_0^2)\int\limits_{-\theta_{in}}^{\theta_{in}}
\frac{ {\rm d}\theta'}{\pi}\lambda(\theta') h(\theta) h(\theta')\,,
\end{eqnarray}
where
\begin{equation}\label{V}
V(\theta)= \sum\limits_{n=0}^2 g_n \sin[n\theta-\Phi(\theta)] - \frac{1-z_0^2}
{z_0^3} \left(g_0+g_1 z_0+g_2 z_0^2 -C(z_0) f_0(\Delta_{K\pi})\right)
h(\theta).
\end{equation}
As discussed in Refs. \cite{AuMa}, \cite{Ca}, for $\Phi(\theta)$ given in
(\ref{Phi1}) and assuming that the physical phase $\delta(\theta)$ is 
sufficiently smooth (more exactly Lipschitz continuous of  order greater than 
1/2 \cite{AuMa}), one can show that (\ref{inteqCTW}) is a Fredholm 
equation.

The minimum of  the left hand side of (\ref{L2gn}) is obtained by inserting the
solution (\ref{gn}) into this relation, taking into account the expression
(\ref{mu}) of $\mu$ and the integral equation (\ref{inteqCTW}). We thus obtain
the inequality
\begin{eqnarray}\label{g0g1g2CTW}
g_0^2 + g_1^2+g_2^2 + \frac{1-z_0^2}{z_0^6} \,\left(g_0+g_1 z_0+g_2 z_0^2
-C(z_0) f_0(\Delta_{K\pi})\right)^2\nonumber\\
+\int\limits_{-\theta_{in}}^{\theta_{in}}\frac
{ {\rm d}\theta} {\pi} \lambda(\theta) V(\theta) \,
\le  \psi^{''}(Q^2)\,,
\end{eqnarray}
where $\lambda(\theta)$ is the solution of the integral equation 
(\ref{inteqCTW}).
The inequality  (\ref{g0g1g2CTW})  defines the allowed domain for the slope and
curvature when both Callan-Treiman relation and Watson theorem are imposed.
We notice that if the Callan-Treiman relation (\ref{CT}) is removed, the 
domain allowed for the coefficients $g_0$, $g_1$ and $g_2$
is given by the inequality
\cite{Ca}
\begin{equation}\label{g0g1g2W}
g_0^2 + g_1^2+g_2^2 + \int\limits_{-\theta_{in}}^{\theta_{in}}\frac
{ {\rm d}\theta} {\pi} \lambda(\theta) V(\theta) \,
\le  \psi^{''}(Q^2)\,,
\end{equation}
where $\lambda(\theta)$ is  the solution of the simpler equation \cite{Ca}
\begin{equation}\label{inteqW}
\sum\limits_{n=0}^2 g_n
\sin[n\theta-\Phi(\theta)]=\lambda(\theta)
-\int\limits_{-\theta_{in}}^{\theta_{in}}\frac{ {\rm d}\theta'} 
{2\pi} \lambda(\theta')\,\frac{\sin[5/2(\theta-\theta')
-\Phi(\theta)+\Phi(\theta')]} {\sin (\theta-\theta')/2}.
\end{equation}

Before ending this section we mention that the alternative inequality
(\ref{L21}) obtained from the dispersion relation (\ref{drpsi1})
leads to similar bounds, which can be obtained by replacing in the r.h.s. of
the inequalities (\ref{g0g1g2}), (\ref{g0g1g2CT}), (\ref{g0g1g2CTW}) and
(\ref{g0g1g2W}) the quantity  $\psi^{''}(Q^2)$ by
$(\psi(Q^2)/Q^2)'+\psi(0)/Q^4$, and the outer function $C(z)$ from (\ref{C}) by
\begin{equation}\label{Ctilde}
\tilde C(z)= C(z)\,\frac{1-z+\beta_Q (1+z)}{2 \sqrt{2}},
\end{equation}
where $\beta_Q$ was defined above (\ref{C}).
Then, instead of (\ref{rel}), we obtain the relations between $g_n$ and the
Taylor coefficients in (\ref{Taylor}) as:
\begin{eqnarray}
g_0&=&   f_0(0)\,0.00249987,  \nonumber \\
g_1&=& f_0(0)\,[ -0.00154365 + 0.0172731 \,\rs^{K\pi}\, ], \label{rel1}\\
g_2&=& f_0(0) [-0.00193761 - 0.0452122 \,\rs^{K\pi}\,  + 0.00651435\, c\,]
\,.\nonumber
\end{eqnarray}
\section{Results and discussion}
The input required for the numerical evaluation of the bounds is represented
by:
\begin{enumerate}
\item
the function $\psi^{''}(Q^2)$ appearing in the right-hand side of the
inequalities (\ref{g0g1g2}), (\ref{g0g1g2CT}), (\ref{g0g1g2CTW}) and
(\ref{g0g1g2W}) 
\item
the ratio $F_K/F_\pi$ appearing the Callan-Treiman relation
(\ref{CT}), 
\item
the phase $\delta (\theta)=\delta_0^{1/2}(t(e^{i\theta}))$
entering the expression (\ref{Phi}) of the function $\Phi$.
\end{enumerate}
\begin{figure}
\vspace{3.em}\hspace{3.3em}\includegraphics[width=11.5cm]{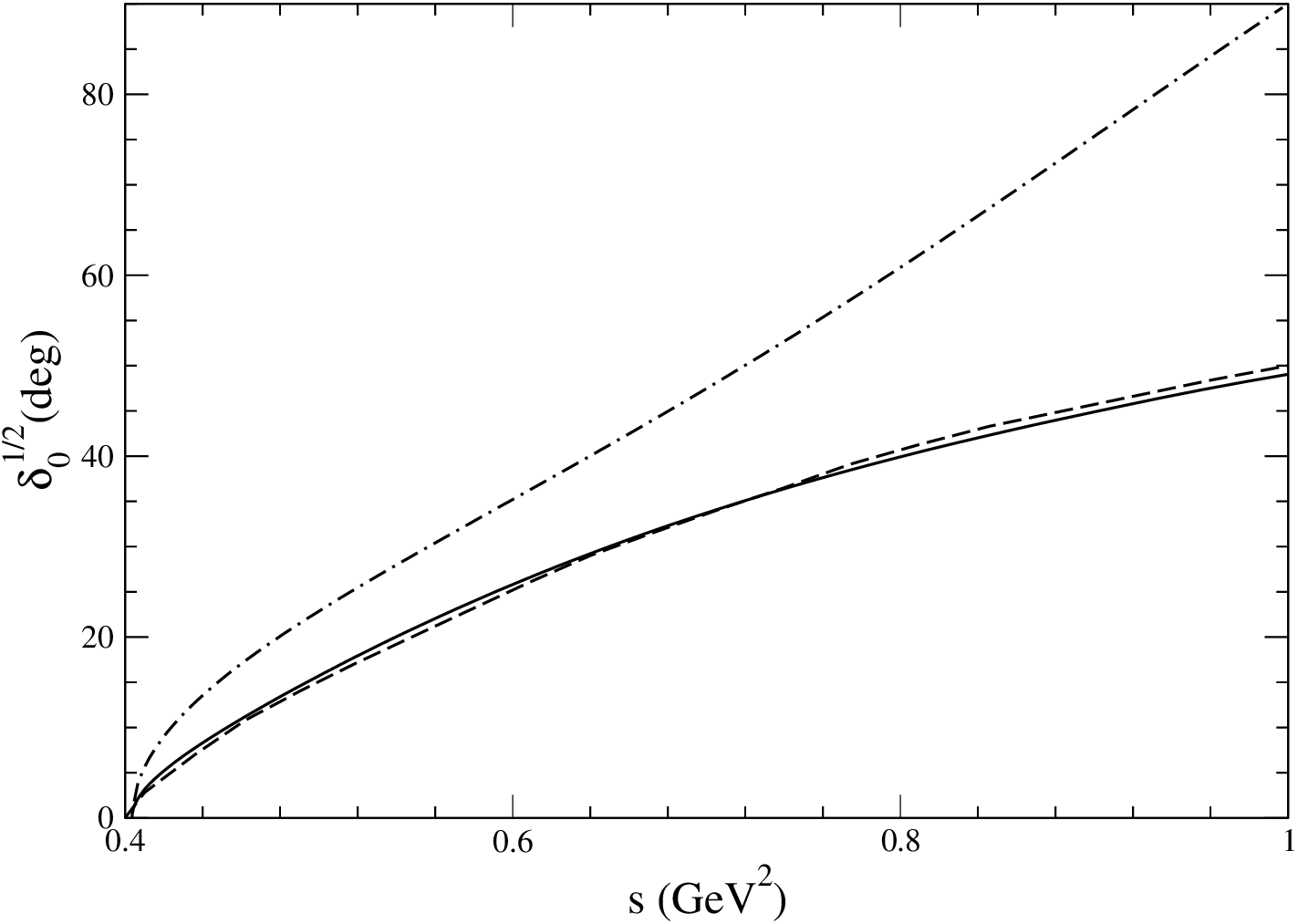}
\caption{\label{fig1:} Phase shift of the $I=1/2$ $S$-wave $K\pi$ amplitude
below 1 GeV. Solid:  phase-shift
$\delta_0^{1/2}(t)$ calculated in \cite{BuDeMo}, used in our analysis;
dashed: one of the curves for $\delta_0^{1/2}(t)$ given in Fig. 1 of
\cite{Jamin_Oller_Pich_2000};  dashed-dot:  parametrization of
$\delta_0^{1/2}(t)$ used in \cite{Ynd}.}
\end{figure}
We calculated $\psi^{''}(Q^2)$ given by the expression (\ref{QCD}) taking
$Q^2=4\, {\rm GeV}^2$ (see Ref.\cite{BoMaRa}),
which is sufficiently large to ensure
the validity of perturbative QCD. At this value of  $Q^2$ the $\alpha_s$
correction in  (\ref{QCD}) is of about 30\%, the $\alpha^2_s$ correction is
less than 9\%,  while the higher twist terms contribute with less than 1\%. The
choice $Q^2=4\, {\rm GeV}^2$ is also convenient since predictions for the light
$m_l$ and strange $m_s$ quark masses  at 2 GeV are now available from 
lattice calculations:
QCDSF-UKQCD Collaboration \cite{QCDSF} predicts $m_s(2\, {\rm GeV})=119(5)(8)
$MeV and $m_l(2\, {\rm GeV})=4.7(2)(3)$ MeV, where the first quoted error is
statistical and the second systematic, while MILC Collaboration \cite{MILC1}
predicts $m_s(2 \,{\rm GeV})=76(0)(3)(7)(0) $MeV and $m_l(2\, {\rm
GeV})=2.8(0)(1)(3)(0)$ MeV, where the errors are from statistics, simulation,
perturbation theory and electromagnetic effects, respectively. As concerns  
the strong
coupling, the  value $\alpha_s (m_\tau^2)$ is now well known from hadronic 
$\tau$ decays
 \cite{ALEPH}, \cite{OPAL}.  The value given in PDG2004, which we used in our 
work, 
is  $\alpha_s (m_\tau^2) = 0.345\pm 0.03$  \cite{PDG2004}, 
slightly larger than the renormalon-based estimates \cite{tau}. We notice that
the strength of the bounds depend in a monotonous way on the magnitude of
$\psi^{''}(Q^2)$, a larger  value of $\psi^{''}(Q^2)$ at a given $Q^2$ leads to
weaker bounds. Therefore, using instead of  $\alpha_s (Q^2)$ at
$Q^2=4\, {\rm GeV}^2$ the value $\alpha_s (m_\tau^2)$ is a conservative 
approximation.
In this way, we calculate $\psi^{''}(4\, {\rm GeV}^2)$ quite accurately
using the running quark masses and the  running coupling at 2 GeV, without
going through the invariant masses and $\Lambda_{QCD}$. Introducing standard 
values of the quark condensates \cite{SVZ}, we obtained 
$\psi''(4\, {\rm GeV}^2)=0.00020$ with the quark masses from \cite{QCDSF},
and $\psi''(4\, {\rm GeV}^2)=0.000079$ with the quark masses from \cite{MILC1}.

The Particle Data Group \cite{PDG2004} reports the value 1.22 for the ratio
$F_K/F_\pi$ appearing the Callan-Treiman relation (\ref{CT}). The pseudoscalar
decay constants were recently  calculated from the lattice \cite{MILC2}, 
with the result  $F_K/F_\pi=1.21 (4) (13)$, where the first 
error is statistical and the second systematic. 
Following Ref. \cite{Jamin_Oller_Pich_2004}, we shall use in our analysis  
$F_K/F_\pi$ in the range 1.21 - 1.23.
\begin{figure}
\vspace{3em}\hspace{3.3em}\includegraphics[width=11.5cm]{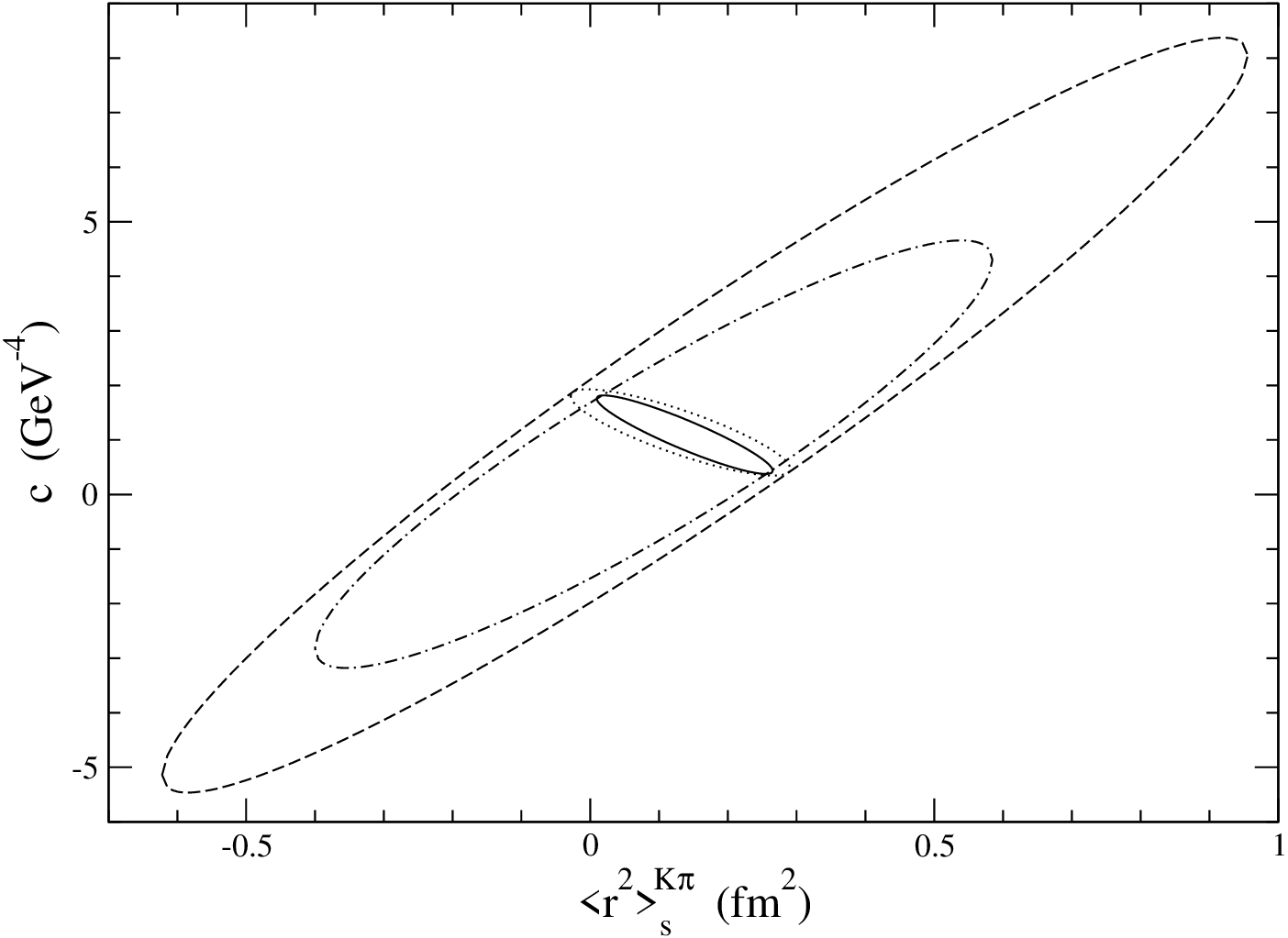}
\caption{ \label{fig2:}  Allowed domains for the $K\pi$ radius squared  and the
curvature  for $f_0(0)=0.976 $ and  $F_K/F_\pi=1.21$; 
dashed: unitarity domain given by Eq. (\ref{g0g1g2});
dashed-dot: domain given by Eq. (\ref{g0g1g2W})
imposing Watson theorem; dotted: domain given by Eq. (\ref{g0g1g2CT}) imposing
Callan-Treiman condition; solid: domain given by Eq.
(\ref{g0g1g2CTW}) imposing Callan-Treiman and Watson theorem.} 
\end{figure}
The last input is provided by the phase shift $\delta_0^{1/2}(t)$ of the
$S$-wave $K\pi$ amplitude along the elastic region. We used in our work the
phase shift derived from a new analysis of $K\pi$ scattering
from Roy-Steiner equation \cite{BuDeMo} 
and the experimental data \cite{PiKexp}. We took the
inelastic threshold at $t_{in}=1 \,{\rm GeV}^2$, since the inelasticity 
is equal to 1 up to this energy. In Fig. \ref{fig1:} we show the phase
$\delta_0^{1/2}(t)$ used in our work, together with  the phase shift
derived in \cite{Jamin_Oller_Pich_2000} and used in the subsequent analysis
\cite{Jamin_Oller_Pich_2002}, \cite{Jamin_Oller_Pich_2004} of the
strangeness changing scalar form factors. For completeness we represent also in
Fig. \ref{fig1:} a parametrization of  $\delta_0^{1/2}(t)$ used in \cite{Ynd}
for a calculation of the radius $\rs^{K\pi}$ based on the single-channel 
Omn\`es formalism.

With the input described above we evaluated the inequalities  (\ref{g0g1g2}),
(\ref{g0g1g2CT}), (\ref{g0g1g2CTW}) and
(\ref{g0g1g2W}),  derived in the previous section. The numerical solution of
the integral equations (\ref{inteqCTW}) and (\ref{inteqW}) was found easily, 
since the coefficients $g_0$, $g_1$ and $g_2$ do not appear in the kernel.
For a fixed value of  $f_0(0)$,  the allowed domain for the slope and curvature
is in all cases the interior of an ellipse in the plane $(\rs^{K\pi},\, c)$. In
order to show the effect produced by each  additional constraint, 
we indicate  in Fig. \ref{fig2:} the domains given by the inequalities
(\ref{g0g1g2}), (\ref{g0g1g2CT}), (\ref{g0g1g2W}) and (\ref{g0g1g2CTW}),
respectively. The curves shown were obtained with  $f_0(0)=0.976$,   the quark
masses from \cite{MILC1} and $F_K/F_\pi=1.21$. One can see that  the most
important effect is produced by the Callan-Treiman condition (\ref{CT}). 
Imposing Watson theorem in addition to the Callan-Treiman
relation leads only to a small reduction of the allowed domain.

\begin{figure}
\vspace{3em}\hspace{3.3em}\includegraphics[width=11.5cm]{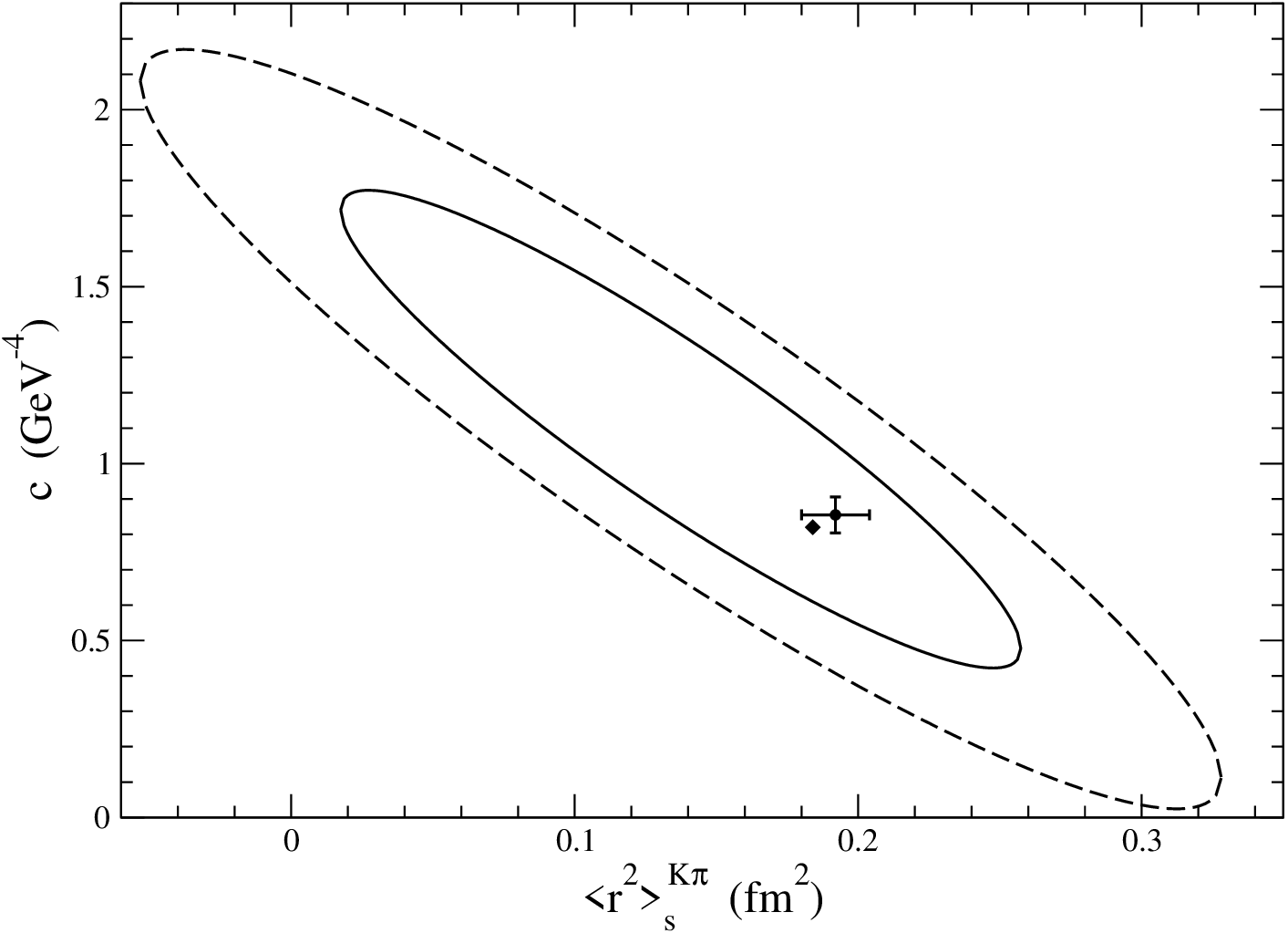}
\caption{\label{fig3:} Allowed domain for the $K\pi$ radius squared   and
curvature, given by Eq. (\ref{g0g1g2CTW})
which incorporates Callan-Treiman relation and Watson theorem. Dashed: values
obtained with $m_s=119$ MeV, $m_u=3$ MeV cf. \cite{QCDSF};
solid: values obtained with $m_s=76$ MeV, $m_u=2.8$ MeV cf. \cite{MILC1}.
The diamond is the determination made in \cite{Jamin_Oller_Pich_2004} for the
same input values $f_0(0)=0.976$ and $F_K/F_\pi=1.21$, while the 
circle with error bars is the final determination in 
\cite{Jamin_Oller_Pich_2004}.}
\end{figure}

In Fig. \ref{fig3:} we indicate the allowed domain given by the inequality
(\ref{g0g1g2CTW}), which takes into account both Callan-Treiman 
condition (\ref{CT}) and Watson theorem (\ref{Watson}), using
for the quark masses the values given in Refs. \cite{MILC1} and \cite{QCDSF},
respectively. As in Fig. \ref{fig2:} we took $f_0(0)=0.976$ and
$F_K/F_\pi=1.21$.  Using the quark masses from \cite{MILC1} we obtain the upper
bound  $\rs^{K\pi} \le 0.26\, {\rm fm}^2$, while with the quark masses from
\cite{QCDSF} we obtain $\rs^{K\pi} \le 0.34\, {\rm fm}^2$. 
In Fig. \ref{fig3:} we indicate also the prediction of the radius and 
curvature  made in \cite{Jamin_Oller_Pich_2004} by means of an
alternative dispersion theory, which gives
$\rs^{K\pi}= 0.192\pm 0.012\, {\rm fm}^2$,   
$c=0.855\pm 0.051\, {\rm GeV}^{-4}$.
These values are situated well inside the allowed domain derived with both
values of the strange quark mass. 
Also, the $\chi$PT value  $\rs^{K\pi}= 0.20\pm 0.05\,
{\rm fm}^2$ derived in  \cite{GL_form_factors} is consistent with the bounds,
while  the value $\rs^{K\pi} =0.31 \pm 0.06\, {\rm fm}^2$
derived in \cite{Ynd} slightly violates the upper bound obtained with the
$s$ quark mass from \cite{MILC1}.

Fig. \ref{fig3:} also predicts upper and lower limits on the curvature $c$ for
each value of  $\rs^{K\pi}$.
For instance, for the ISTRA value \cite{Yushchenko_2003}:
$\rs^{K^\pm\pi}=0.235\,\mbox{fm}^2 $, and the curvature $c$ must be in 
the range from  0.41 to 0.78 GeV$^{-4}$, while for the KTeV value
\cite{KTeV_form_factors_2004} : $\rs^{K_L\pi}= 0.165\,\mbox{fm}^2$, and 
the curvature must be larger than 0.67 and smaller than 1.23 GeV$^{-4}$.
Thus, the second order term in the Taylor expansion (\ref{Taylor})
seems to be important and can not be neglected in experimental fits. The
allowed range of $c$   given in Fig. \ref{fig3:} for each value of the slope
may be useful as an additional constraint in the representation of the
experimental data.
\begin{figure}
\vspace{3em}\hspace{3.3em}\includegraphics[width=11.5cm]{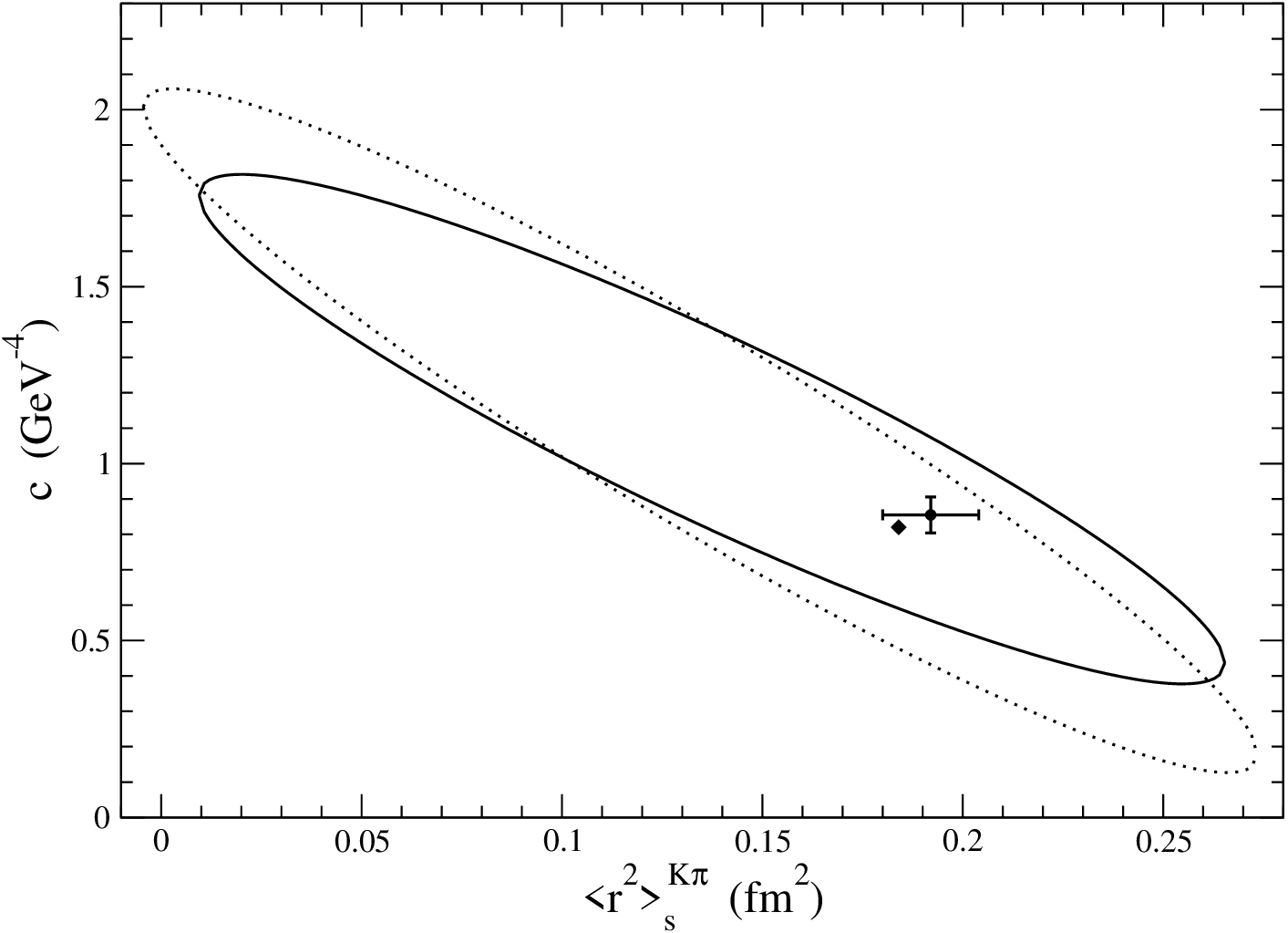}
\caption{\label{fig4:} Allowed domain for the $K\pi$ radius squared and
curvature given by Eq.(\ref{g0g1g2CTW})  using $m_s=76$ MeV, 
$m_u=2.8$ MeV: solid: domain obtained
with the dispersion relation (\ref{drpsi}); dotted: domain obtained with the
dispersion relation (\ref{drpsi1}).
The points are the determination made in \cite{Jamin_Oller_Pich_2004}.
}\end{figure}
Up to now we indicated the allowed domains  obtained with the form
(\ref{drpsi}) of the dispersion relation. 
The relation (\ref{drpsi1}) leads to similar results. In this case, in the 
r.h.s. of the inequalities (\ref{g0g1g2}), (\ref{g0g1g2CT}), (\ref{g0g1g2W}) 
and (\ref{g0g1g2CTW}) appears the quantity $(\psi(Q^2)/Q^2)'+\psi(0)/Q^4$. 
We evaluated this quantity at $Q^2=\,4{\rm GeV}^2$ 
using the QCD expression (\ref{QCD1}) and the inequality (\ref{psi0}), 
which leads to the values 0.00022 with the quark masses from \cite{MILC1},  
and 0.00038 with the quark masses from \cite{QCDSF}. In Fig. \ref{fig4:} we
present for comparison the allowed domain for the radius and curvature obtained
with the two forms of the dispersion relation discussed above.

It is of interest to combine the bounds on the Taylor coefficients of $f_0(t)$
derived in the present work with the predictions of  $\chi$PT.
The expression of the scalar form factor at order $p^6$ in $\chi$PT is
\cite{Bijnens_Talavera_2003}
\begin{equation}\label{chPT}
f_0(t)=F_+(0)+\bar\Delta(t) +\frac{F_K/F_\pi-1}{\Delta_{K\pi}}\,t
+\frac{8}{F_\pi^4} (2 C_{12}^r +C_{34}^r)\Sigma_{K\pi} t - \frac{8}{F_\pi^4}
C_{12}^r t^2,
\end{equation}
where
\begin{eqnarray}\label{Fp0}
F_+(0)&=&1-0.008 -\frac{8}{F_\pi^4} (C_{12}^r +C_{34}^r)\Delta_{K\pi}^2
, \nonumber\\
\bar\Delta(t) &=& -0.259 t + 0.84 t^2 + 1.291 t^3,
\end{eqnarray}
where the chiral constants are taken at the scale $\mu=M_\rho$ 
\cite{Bijnens_Talavera_2003}, \cite{Jamin_Oller_Pich_2004}.
The inequalities derived in the previous section can be expressed
as an allowed domain for the chiral constants $C_{12}^r$
and $C_{34}^r$. As in  Ref.\cite{Jamin_Oller_Pich_2004} we use as independent
parameters $C_{12}^r$ and the sum $C_{12}^r +C_{34}^r$.  
For  $Q^2=4\, {\rm GeV}^2$, the first three Taylor coefficients $g_n$ defined 
in (\ref{gTaylor}) are expressed in terms of these parameters as
\begin{eqnarray}
g_0&=&   0.00163 - 9.23\, (C_{12}^r +C_{34}^r), \nonumber \\
 g_1&=&  -0.000104 + 77.2\, C_{12}^r + 87.84\,  (C_{12}^r +C_{34}^r) \,.
\label{relC} \\
g_2&=&-0.00226 - 713.3 \,C_{12}^r - 241.93 \, (C_{12}^r +C_{34}^r)
\,.\nonumber
\end{eqnarray}
\begin{figure}
\vspace{3.0em}\hspace{3.3em}
\includegraphics[width=11.5cm]{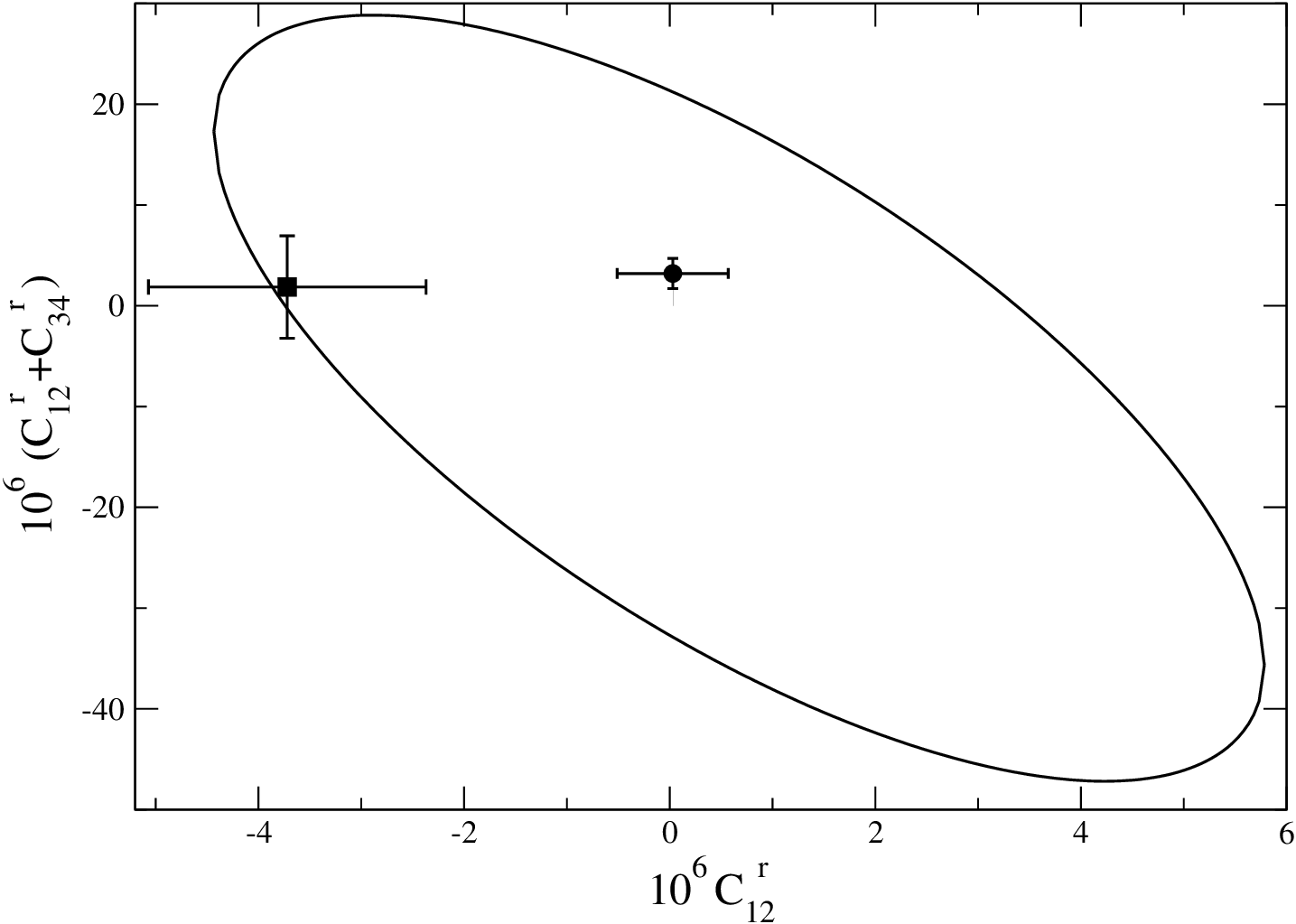}
\caption{ \label{fig5:}  Allowed domains for the low energy constants
$C_{12}^r$ and $C_{12}^r+C_{34}^r$ using Eq. (\ref{g0g1g2CTW}), 
with the quark masses from \cite{MILC1} and $F_K/F_\pi=1.21$.
The circle with errors is the determination in \cite{Jamin_Oller_Pich_2004}, 
the square is
the prediction in \cite{CiEcEi}, where the quoted errors represent
the variation with the scale $\mu$ in the range from $M_\eta$ to  1 GeV.}
\end{figure}
Then inequality (\ref{g0g1g2CTW}) gives the allowed domain  shown in 
Fig.\ref{fig5:}, where we indicate also the result derived in 
\cite{Jamin_Oller_Pich_2004}  and  
the recent determination of the chiral low-energy constants in \cite{CiEcEi}.

From Figs. \ref{fig3:}-\ref{fig5:} it is seen that the predictions made in
\cite{Jamin_Oller_Pich_2004} are much more precise than the unitarity bounds
derived by us, even with the implementation of Callan-Treiman and Watson
theorem. It is of interest to see what is the additional input used in
\cite{Jamin_Oller_Pich_2004}, responsible for the very small uncertainties
quoted for their results. In Ref. \cite{Jamin_Oller_Pich_2002} the authors 
solve the coupled channel unitarity equations for the 
$K\pi$, $K\eta'$ and $K\eta$ form
factors, using  the chiral resonance model and various $K$-matrix
parametrizations of the scattering amplitudes. By construction, 
the Watson theorem is satisfied by the
$K\pi$ form factor $f_0(t)$, while the two integration constants entering the
solution of the coupled channel equations are fixed by the values of  $f_0(t)$
at $t=0$ and $t=\Delta_{K\pi}$ cf. Callan-Treiman relation (\ref{CT}).
Therefore, $f_0(t)$ calculated in \cite{Jamin_Oller_Pich_2002} satisfies the
same low energy constraints which we imposed. However, in our formalism the 
form factor is
constrained above the inelastic threshold  only by the $L^2$ condition
(\ref{L2}) (or (\ref{L21})). This puts a weak restriction on the modulus 
$|F(t)|$, which is 
allowed to increase like  $t$
at asymptotic energies, while  in \cite{Jamin_Oller_Pich_2002} the  $1/t$ 
asymptotic decrease predicted by 
perturbative QCD is imposed.  Analyticity is also implemented in a different 
way in the two approaches: in 
our formalism it is expressed by the relations (\ref{g}) and (\ref{gTaylor}),
where the real
coefficients $g_n$ obey only the constraints that were explicitly 
incorporated. So, 
we work with the most general class of functions which satifies these 
constraints. 
On the other hand, in  \cite{Jamin_Oller_Pich_2002} the form factor
is expressed by a dispersion relation with the spectral function related to 
the meson - meson
transition amplitudes, described by several specific parametrizations.
Choosing a
restricted class of functions leads to a smaller error, but may introduce a
bias, so the errors given in 
\cite{Jamin_Oller_Pich_2002} might be underestimated.

Of course, the information provided by the inelastic channels is important.
The contribution of other two-particle states can be implemented in a 
rigorous way
in the present formalism, as shown in \cite{Ca} (see also \cite{CaLeNe}). 
The higher states in the unitarity
relation (\ref{unit}) involve the modulus squared of the $K\eta$
and $K\eta'$  form factors  above the corresponding unitarity thresholds. 
Their unphysical
cuts can be eliminated by exploiting  Watson theorem with a generalized Omn\`es
formalism \cite{Ca}.  We shall investigate this problem in a future work.
Other generalization, easily implementable, is the introduction of higher
derivatives at zero momentum transfer, which are related to the chiral
constants of $\chi$PT.

We mention finally that the inequalities derived above can be used also in the
opposite way, for bounding the quark masses using the information on the low
energy expansion parameters of the strangeness changing scalar form factors.
This problem was investigated for instance in \cite{LeRaTa} and \cite{JaOlPi}.
Our approach offers the possibility to include simultaneously various pieces of
information, like Callan-Treiman constraint (\ref{CT}) and  $K\pi$ phase shift
$\delta_0^{1/2}$ through Watson theorem. For instance, using the predictions of
the radius and the curvature made in \cite{Jamin_Oller_Pich_2004} we obtain the
lower bound $m_s (2\ {\rm GeV})-m_u (2\ {\rm GeV})>38.6\,{\rm MeV}$.
The prediction made in \cite{JaOlPi} based on a QCD sum rule is tighter since
the authors use 
the spectral function of the two point function $\psi$ calculated from the 
values of the $K\pi$,  $K\eta$ and $K\eta'$
form factors on the unitarity cut.

\section{Conclusions}
We have derived and evaluated unitarity bounds on the slope and curvature of
the scalar strangeness changing form factor $f_0(t)$. The method incorporates
in an exact mathematical way low energy theorems like Callan-Treiman condition
and Watson final state interaction theorem. Our results indicate that the
curvature can not be neglected in the representation of the experimental data
on $K_{l3}$ decay. Using the calculation of the form factor to two loops in
$\chi$PT, we expressed our results as an allowed domain for the chiral
constants $C_{12}^r$ and $C_{12}^r+C_{34}^r$. Our bounds confirm the values
obtained recently in an coupled channel dispersion approach, but are much
weaker (however  the
uncertainties given in \cite{Jamin_Oller_Pich_2004} might be underestimated).
Generalizations of the method, including the $K\eta$ and $K\eta'$ form factors
in the unitarity sum for the spectral function, 
and their higher derivatives  at the origin, can increase the predictive power
of the present formalism.

\subsection*{Acknowledgments} We are indebted to G. Colangelo, J. Gasser, M.
Knecht, L. Lellouch and H. Leutwyler for interesting discussions. 
UMR 6207 is Unit\'e Mixte de Recherche du CNRS and of Universit\'es
Aix-Marseille I and Aix-Marseille II and of Universit\'e du Sud
Toulon-Var, laboratoire affili\'e \`a la FRUMAN.
I.C. thanks the Centre de Physique Th\'eorique de Marseille for hospitality. 
This work was realized in the frame of the Cooperation Agreement 
between CNRS and the Romanian Academy 
(Project CPT-Marseille - NIPNE Bucharest) and was supported by
MEC-Romania in the frame of CERES Program, Contract C4-123.


\begin{thebibliography}{99}

\bibitem{KTeV_form_factors_2004}
KTeV Collaboration, T. Alexopoulos, et al., hep-ex/0406003.

\bibitem{Donaldson_1974}
G. Donaldson, et al., Phys. Rev. D 9 (1974) 2960.

\bibitem{Yushchenko_2003}
O.P. Yushchenko, et al., Phys. Lett. B 581 (2004) 31, hep-ex/0312004.

\bibitem{AdGa} M. Ademollo, R. Gatto, Phys. Rev. Lett. 13 (1964) 264.

\bibitem{Dashen_Weinstein}
C.G. Callan, S.B. Treiman, Phys. Rev. Lett. 16 (1966) 153;\\
R.F. Dashen, M. Weinstein, Phys. Rev. Lett. 22  (1969) 1337.

\bibitem{GL_form_factors}
J. Gasser,  H. Leutwyler, Nucl. Phys. B 250 (1985) 517.

\bibitem{Post_Schilcher_2002}
P. Post, K. Schilcher, Eur. Phys. J. C 25 (2002) 427, hep-ph/0112352.

\bibitem{Bijnens_Talavera_2003}
J. Bijnens, P. Talavera, Nucl. Phys. B 669 (2003) 341, hep-ph/0303103.

\bibitem{GL_1985}
J. Gasser, H. Leutwyler, Nucl. Phys. B 250 (1985) 465, eq. (11.6).

\bibitem{Okubo} S. Okubo, Phys. Rev. D 3 (1971) 2807; ibid D 4 (1971) 725.

\bibitem{LiPa} L.F. Li, H. Pagels, Phys. Rev. D 3 (1971) 2191.

\bibitem{MaOk} V. Mathur, S. Okubo, Phys. Rev. D 1 (1971) 3468.

\bibitem{Bour} C. Bourrely, Nucl. Phys. B 43 (1972) 434; ibid B
53 (1973) 289.

\bibitem{AuMa} G. Auberson, G. Mahoux, F.R.A. Somao, Nucl. Phys. B 98
(1975) 204.

\bibitem{BNRY} C. Becchi, S. Narison, E. de Rafael, F.J. Yndurain, Z. Phys.
C 8 (1981) 335.

\bibitem{BoMaRa} C. Bourrely, B. Machet, E. de Rafael, Nucl. Phys. B
189 (1981) 157.

\bibitem{Jamin_Oller_Pich_2002}
M. Jamin, J.A. Oller, A. Pich, Nucl. Phys. B 622 (2002) 279, hep-ph/0110193.

\bibitem{Jamin_Oller_Pich_2000}  M. Jamin, J.A. Oller, A. Pich, Nucl. Phys.
B 587 (2000) 331,\\ hep-ph/0006045 .

\bibitem{Jamin_Oller_Pich_2004}
M. Jamin, J.A. Oller, A. Pich, JHEP 0402 (2004) 047, hep-ph/0401080.

\bibitem{RadCor}  V. Cirigliano, M. Knecht,  H. Neufeld, H. Rupertsberger, 
P. Talavera, 
Eur. Phys. J. C23 (2002) 121,  hep-ph/0110153.
\bibitem{GoKa} S.G. Gorishny, A.L. Kataev, S.A. Larin, R.L. Surguladze,
Mod. Phys. Lett. A 5 (1990) 2703.

\bibitem{Chet} K.G. Chetyrkin, Phys. Lett. B 390 (1997) 309.

\bibitem{QCDSF} QCDSF-UKQCD Collaboration, M. G\"ockeler, et al., 
hep-ph/0409312.

\bibitem{MILC1}
MILC Collaboration, C. Aubin. et al., Phys.Rev.D 70 (2004)  031504.

\bibitem{MILC2}
MILC Collaboration, C.Aubin, et al.,
Phys. Rev. D 70 (2004) 114501, hep-lat/0407028.

\bibitem{Watson} K.M. Watson, Phys. Rev. 95 (1954) 228.

\bibitem{Ca} I. Caprini, Eur. Phys. J .C 13 (2000) 471, hep-ph/9907227.

\bibitem{Omnes} R. Omn\`es, Nuovo Cim. 8 (1958) 316.

\bibitem{Musk} N.I. Muskchelishvili, Singular Integral Equations, Noordhoff,
Groningen, 1953.

\bibitem{SVZ} M.A. Shifman, A.I. Vainstein, V.I. Zakharov, Nucl. Phys.
B 147 (1979) 385.

\bibitem{ALEPH} R. Barate et al., (ALEPH Collaboration), Eur. Phys. J. C4 
(1998) 409.

\bibitem{OPAL} K. Ackerstaff et al., (OPAL Collaboration), Eur. Phys. J. C7 
(1999) 571.

\bibitem{PDG2004} S. Eidelman, et al., Phys. Lett. B 592 (2004) 1.
\bibitem{tau} P. Ball, M. Beneke, V. M. Braun,  Nucl. Phys. B 452 (1995)
563;\\
M. Neubert, Nucl. Phys. B 463 (1996) 511.


\bibitem{BuDeMo} P. Buettiker, S. Descotes-Genon, B. Moussallam, Eur. Phys. J.
C 33 (2004) 409, hep-ph/0310283.

\bibitem{PiKexp} P. Easterbrook, et al., Nucl, Phys. B 133 (1978) 490;\\
D. Aston, et al., Nucl. Phys. B 296 (1988) 493.
\bibitem{Ynd} F. Yndurain, Phys. Lett. B 579 (2004) 99; Erratum: B
586 (2004) 439.

\bibitem{CiEcEi} V. Cirigliano, G. Ecker, M. Eidem\"uller, R. Kaiser, A. Pich,
J. Portol\'es, The $<SPP>$ Green function 
and SU(3) breaking in $K_{l3}$ decays,  hep-ph/0503108.

\bibitem{CaLeNe} I. Caprini, L. Lellouch, M. Neubert, Nucl. Phys. B 530
(1998) 153.

\bibitem{LeRaTa} L. Lellouch, E. de Rafael, J. Taron, Phys. Lett. B 414
(1997) 195.

\bibitem{JaOlPi} M. Jamin, J. A. Oller, A. Pich, Eur. Phys. J. C 24 (2002)
237.

\end{thebibliography}
\end{document}